\documentclass[aps,prl,reprint,superscriptaddress]{revtex4-2}
\usepackage{epsfig}
\usepackage{graphicx}
\usepackage[caption=false]{subfig}
\usepackage{bm}
\usepackage{color}
\usepackage{float}
\usepackage{mathtools}
\usepackage{amsmath}
\usepackage{comment}
\usepackage{textcomp}
\usepackage[normalem]{ulem}
\usepackage{makecell}

\newcommand{\blue}{\textcolor{black}}
\newcommand{\red}{\textcolor{black}}

\begin{document}

	\title{\red{Nuclear} spin coherence in superconducting Nb$_3$Sn}
	
	\author{Gan Zhai}
	\altaffiliation{ganzhai2025@u.northwestern.edu}
	\affiliation{Department of Physics and Astronomy, Northwestern University, Evanston IL 60208, USA}
	\author{William P. Halperin}
	\altaffiliation{w-halperin@northwestern.edu}
	\affiliation{Department of Physics and Astronomy, Northwestern University, Evanston IL 60208, USA}
	\author{Arneil P. Reyes}
	\affiliation{National High Magnetic Field Laboratory, Tallahassee FL 32310, USA}
	\author{Sam Posen}
	\affiliation{Fermi National Accelerator Laboratory, Batavia IL 60510, USA}
	\author{Chiara Tarantini}
	\affiliation{National High Magnetic Field Laboratory,  Applied Superconductivity Center, Florida State 			University, Tallahassee FL 32310, USA}
	\author{Manish Mandal}
	\affiliation{National High Magnetic Field Laboratory,  Applied Superconductivity Center, Florida State 			University, Tallahassee FL 32310, USA}	
	\author{David C. Larbalestier}
	\affiliation{National High Magnetic Field Laboratory,  Applied Superconductivity Center, Florida State 			University, Tallahassee FL 32310, USA}
	
	\date{\today}
	
	\begin{abstract}
	 We have investigated the normal  and superconducting states of the technologically important compound Nb$_3$Sn using $^{93}$Nb nuclear magnetic resonance.  From spin-lattice relaxation we find strong suppression of the zero-temperature superconducting order parameter by magnetic field. \red{Additionally} we have identified an anomalously large  electron-nuclear exchange interaction from spin-spin relaxation measurements,  an order of magnitude beyond that of the nuclear dipolar \red{coupling.  This RKKY interaction  evolves from  normal to superconducting states, becoming essentially Lorentzian  in the low temperature limit.}
	\end{abstract}

	\maketitle
	
	\section{Introduction}
	Nb$_3$Sn is a superconductor with high critical temperature $T_c (\sim18\,$K) and high critical field $H_{c2} (\sim30\,$T).~\cite{Zho.11b, Zho.11,Man.25} Its unique physical properties have led to applications in superconducting magnet technology, especially  for achieving very high magnetic fields. However, little is known about this compound from the microscopic perspective provided by nuclear magnetic resonance, $^{93}$Nb (NMR). In our previous brief work, we surveyed the $^{93}$Nb NMR properties of Nb$_3$Sn in both normal and superconducting states.~\cite{Gan.24} Here we report detailed measurements of the spin-lattice, {\it i.e.}\,longitudinal, relaxation rate ($T_1^{-1}$) at low temperatures \red{enabling measurement of} the low temperature superconducting order parameter and its suppression by magnetic field and the  temperature dependence of the spin-spin, {\it i.e.}\,transverse, relaxation rate ($T_2^{-1}$), revealing  high  coherence attributable to an RKKY interaction between nuclear spins.  Previous NMR studies of superconducting compounds, most notably in high-$T_c$ superconductors,\cite{Rec.97, Bac.98}   have related $T_2$ to vortex dynamics. This is also possible for Nb$_3$Sn owing to its high nuclear spin coherence. 
	
	The  measure of nuclear spin coherence is expressed by the relaxation time $T_2$ usually dominated by the direct nuclear dipole-dipole interaction. \red{In metallic materials  nuclear spins are also coupled  through an exchange interaction via conduction electrons.  We discovered that the  $^{93}$Nb spins in Nb$_3$Sn  have an anomalously large Ruderman-Kittel indirect exchange interaction (RKKY)\cite{Rud.54} that reduces the nuclear dipole-dipole interaction leading to substantial narrowing of the NMR spectrum. This can be described as being similar to  extreme motional narrowing in liquids.\cite{And.53,Sli.13} } A prototypical example in a metal is  $^{195}$Pt NMR in pure platinum.\cite{Wal.62} We discovered that Nb$_3$Sn is in the same class of such systems \red{but in this case resulting in} an unusually long  $T_2$, a factor of 20 as compared with the direct dipole-dipole interaction.  \red{Since niobium has a single, non-zero spin isotope, $^{93}$Nb ($I=9/2$), the indirect interaction is energy conserving, a requirement for the narrowing phenomenon to exist. We find that the superconducting state significantly modifies this interaction.}
	
	\section{Sample preparation and characterization}
\begin{figure}
		\includegraphics[width=0.5\linewidth]{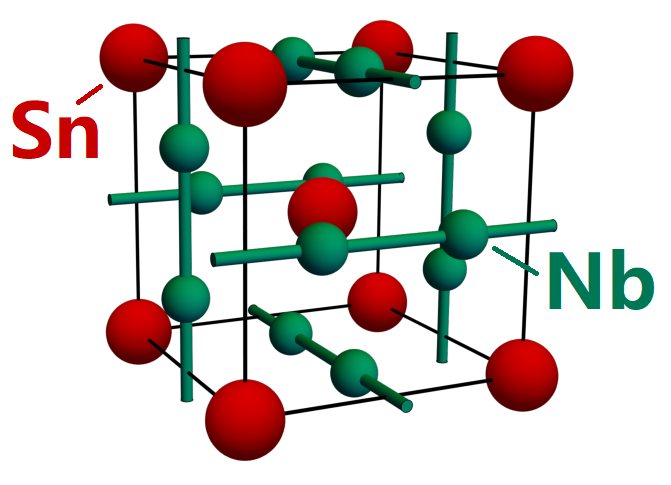}
		\caption{\label{Crystal} Crystal structure of stoichiometric Nb$_3$Sn.}
	\end{figure}

	The high-quality powder sample was produced in the Applied Superconductivity Center of the National High Magnetic Field Laboratory following high-energy ball milling, initial cold isostatic press densification, and final hot isostatic press reaction and densification~\cite{Zho.11}. In this work we have two samples with slightly different preparation. The parameters of the two samples are listed in Table\,~\ref{table1}.  According to X-ray diffraction (XRD) characterization, Nb$_3$Sn has a typical cubic A15 crystal structure, shown in Fig.~\ref{Crystal}. The Sn atoms sit on the corners and the body-centered positions of a unit cell, while the Nb atoms occupy three orthogonal chains on the faces of the cube.
	
	%%%%%%%%%%%%%%%%%%%%%     Figure 1    %%%%%%%%%%%%%%%%%%%%%%%%%
	
	\begin{table}
		\centering
		\renewcommand{\arraystretch}{2}
		\begin{tabular}{|c|c|c|c|c|c|}
			\hline
			Sample & Composition & \makecell{Heat \\ Treatment} & $H_{c2}(0)$ & $T_c$ at 0 field & Weight\\
			\hline
			No.1 & 27\%\,Sn & 1800\,\textdegree{}C & 28.9\,T & 17.4\,K & 20\,mg\\
			\hline
			No.2 & 25\%\,Sn & 1600\,\textdegree{}C & 30.2\,T & 18.0\,K & 220\,mg\\
			\hline
		\end{tabular}
		\renewcommand{\arraystretch}{1}
		\caption{A summary of Nb$_3$Sn powder samples used for NMR measurements; also Supplementary Materials.}
		\label{table1} 
	\end{table}

	\section{Experiment}
	The NMR experiments were performed at the National High Magnetic Field Laboratory with magnetic fields varying from 3.7\,T to 15\,T. The superconducting transition temperature $T_c$ at different fields was identified from heat capacity and resistance measurements consistent with  detuning of the NMR coil at onset of superconductivity.  All measurements were conducted on the central transition of the quadrupolar split spectrum of $^{93}$Nb ($I=9/2$) having eight quadrupolar satellites.(Supplementary Materials).
	
	The $T_2^{-1}$ was measured with a $90^\circ-180^\circ$ pulse sequence, known as the Hahn echo~\cite{Hah.50}, with varying time interval $\tau$ between the two pulses. The $T_2^{-1}$ was obtained by fitting the recovery curve as discussed later. The $T_1^{-1}$ was measured using a $90^\circ-\tau-90^\circ-\tau_0-180^\circ$ pulse sequence, where the Hahn echo was acquired with a fixed $\tau_0$.
	
	\section{Results}
	\subsection{Longitudinal Relaxation}

	The longitudinal relaxation rate $T_1^{-1}$ was measured at the central transition in the superconducting state at 3.7\,T, 7.5\,T, 11.3\,T and 15.0\,T respectively in Sample 1. It was found to be the same at satellite frequencies. The $T_1^{-1}$ was obtained by fitting the recovery curve $M(t)$ to a stretched "Master equation"~\cite{Sut.98} with $\beta \sim 0.9$ for the four pairs of quadrupolar satellites in addition to the central transition:
	
	\begin{equation}
	\begin{split}
		M(t)=M_0 - &M_0[\frac{1}{165}\mathrm{exp}(-t/T_1)^\beta+\frac{24}{715}\mathrm{exp}(-6t/T_1)^\beta\\
		&+\frac{6}{65}\mathrm{exp}(-15t/T_1)^\beta+\frac{1568}{7293}\mathrm{exp}(-28t/T_1)^\beta\\
		&+\frac{7938}{12155}\mathrm{exp}(-45t/T_1)^\beta]
	\end{split}
	\label{Master}
	\end{equation}
	
	%===============figure T1================%
	
	\begin{figure}[htbp]
		\includegraphics[width=\linewidth]{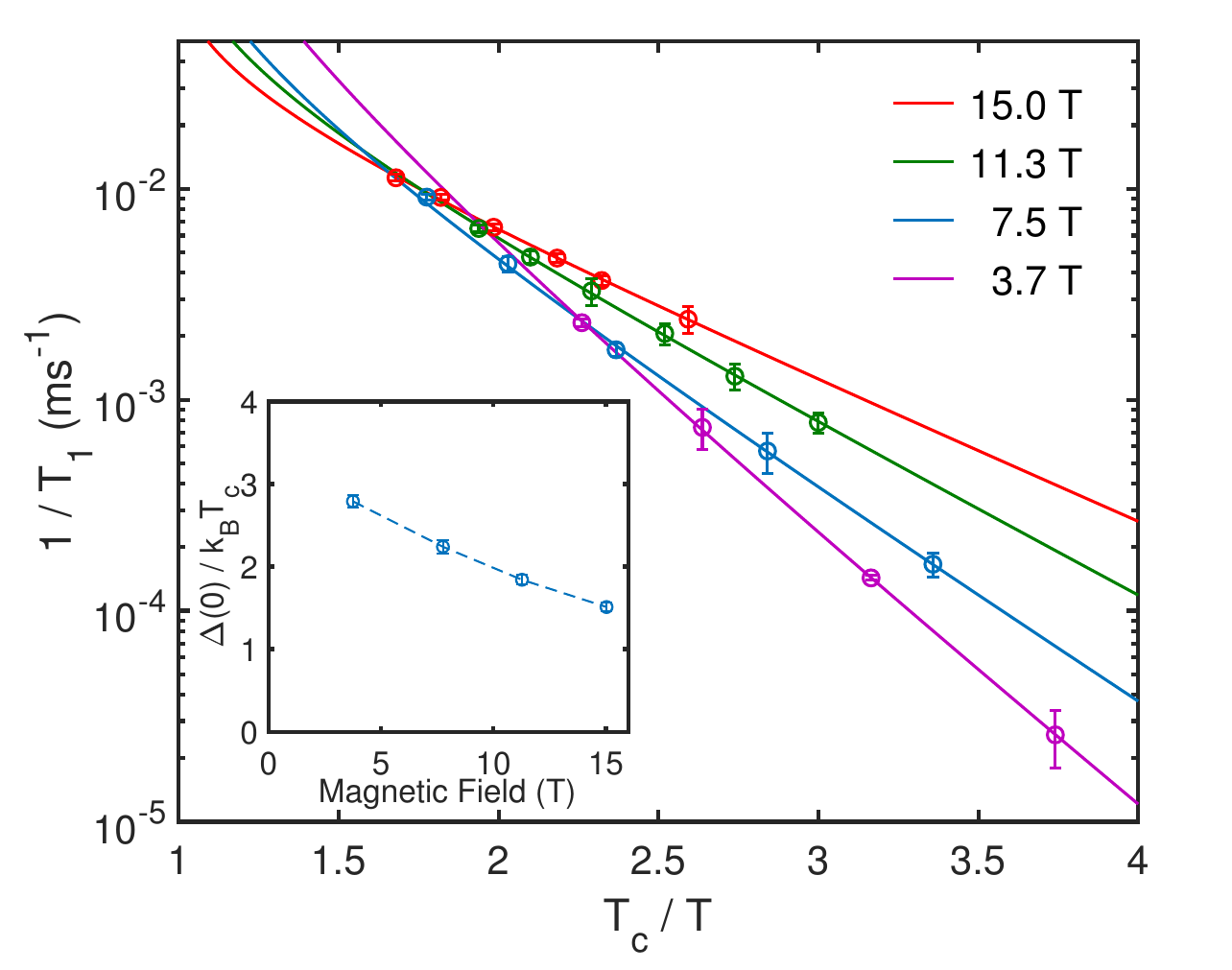}
		\caption{\label{T1} $^{93}$Nb longitudinal relaxation rate $T_1^{-1}$ in the superconducting state of the Nb$_3$Sn powder (Sample 1.). The lines are the fitting results using Eq.~\ref{T1vsDelta} and Eq.~\ref{DeltavsT} at different magnetic fields. The slope approximation in the low temperature region represents the average energy gap at zero temperature ($\Delta(0)$). Inset: $\Delta(0)/k_B T_c$ vs. external field where $\Delta(0)$ comes from the fitting, the dashed line is a guide to the eye.}
	\end{figure}

	The temperature dependence of $T_1^{-1}$ in the low temperature region is plotted in Fig.~\ref{T1}. The data are fitted by Eq.~\ref{T1vsDelta}, which identifies the dependence of $T_1^{-1}$ on the superconducting energy gap $\Delta(T)$, and Eq.~\ref{DeltavsT}, illustrating the temperature dependence of the gap with strong-coupling taken into account~\cite{Hal.90}:
	
	\begin{equation}
		T_1^{-1} \propto \mathrm{exp}(-\Delta(T)/k_BT)
		\label{T1vsDelta}
	\end{equation}
	
	\begin{equation}
		\Delta(T) = \Delta (0) \mathrm{tanh} \{ \frac{\pi k_B T_c}{\Delta(0)} [\frac{2}{3} (\frac{T_c}{T}-1) \frac{\Delta C}{C}]^{1/2} \}
		\label{DeltavsT},
	\end{equation}
	
	\noindent where the heat capacity jump  $\Delta C/C$ at $T_c$ incorporates  strong-coupling in a BCS energy gap. 
	
	The energy gap in the zero temperature limit $\Delta(0)$ was obtained, from the temperature dependence of $T_1^{-1}$ taking $\Delta C/C$ from heat capacity measurements. This is plotted in the inset of Fig.~\ref{T1} as a function of magnetic field. We observe a clear decrease in the energy gap as a function of the external magnetic field, indicating  suppression of the order parameter by field. Similar suppression behavior has been reported in V$_3$Si from STM measurements.\cite{Din.23} The $\Delta(0)$ at 15\,T is even lower than the BCS weak coupling value $1.76\,k_BT_c$, possibly associated with quasiparticle excitations in the vortex cores.\cite{Fet.68}
	
	\subsection{Transverse Relaxation}

	%%%%%%%%%%%%%%%%%%%%%     Figure T2    %%%%%%%%%%%%%%%%%%%%%%%%%
	
	\begin{figure}[htbp]
		\includegraphics[width=\linewidth]{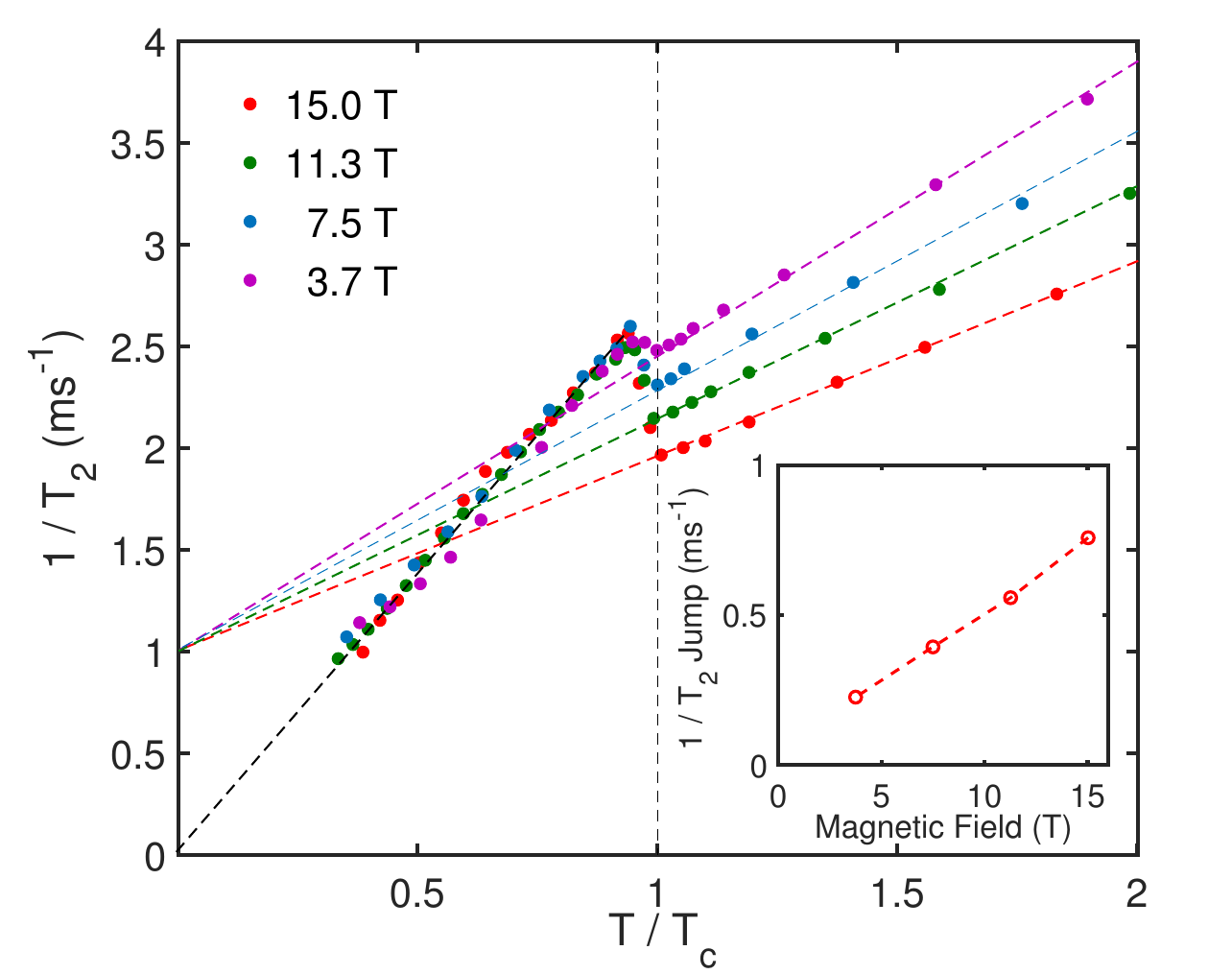}
		\caption{\label{T2} Temperature dependence of $^{93}$Nb transverse relaxation rate $T_2^{-1}$ of the Nb$_3$Sn powder in various fields (Sample 1.). Plotting $T_2^{-1}$ versus temperature ($T/T_c$) and magnetic field clearly shows  common aspects in both the superconducting and normal states in this sample. The dashed lines in the normal state are linear fits that are magnetic field independent in contrast to the superconducting state where the slopes of the lines are a guide to the eye suggesting proportionality to the field since $T_c \propto H$. The extrapolations to $T=0$ are notable.}
	\end{figure}

	The transverse relaxation rate $T_2^{-1}$ was  measured at the central transition in both the normal and superconducting states at various fields.  For Sample 1, this is shown in Fig.~\ref{T2}. The $T_2^{-1}$ was obtained from a stretched exponential fit to measured relaxation curves. In Fig.~\ref{T2}, $T_2^{-1}$ exhibits a linear dependence on temperature in the normal state, despite the fact that the nuclear dipole-dipole interaction is temperature-independent. However, in addition to the dipole-dipole interaction, $T_2^{-1}$ is also constrained by the lifetime of the excited nuclear spin, referred to as the "Redfield effect"~\cite{Red.57,Sli.13}, which explicitly takes into account the dependence of $T_2^{-1}$ on $T_1^{-1}$. From the Korringa law, $T_1^{-1}$ \red{has linear temperature dependence} in the normal state, consequently this linear behavior of $T_{2meas}^{-1}$ strongly suggests a Redfield effect expressed as,
	
	\begin{equation}
		T_{2meas}^{-1} = T_{2}^{-1} + \blue{\kappa \cdot T_{1}^{-1}}
		\label{Redfield}
	\end{equation}
	
	\noindent where $T_{2meas}^{-1}$ is the measured echo response and \blue{$\kappa \cdot T_{1}^{-1}$} represents the Redfield term and $\kappa$ is a constant of order one. 
	
	In  the normal state, $T_2^{-1}$, Fig.~\ref{T2}, is field independent since  $T_c \propto H$ in this range of field (Supplementary Materials), consistent with expectation of  magnetic-field-independence of the dipole-dipole interaction as well as of the density of electronic states that determine $T_1^{-1}$.  However, in the superconducting state, there is clearly a jump of $T_2^{-1}$ at $T_c$, shown in Fig.~\ref{T2}. The magnitude of the jump is plotted in the inset of Fig.~\ref{T2}. The fact that the jump scales with the external field is a clear indication of vortex dynamics since the vortex density is proportional to the field. In the superconducting state, we find a common  linear dependence of $T_2^{-1}$ on temperature $T/T_c$(H) in Fig.~\ref{T2} which means that it  must scale with magnetic field.  Consequently, we  also attribute this field dependence  to vortex dynamics, similar to what is shown in the inset. 
	%%%%%%%%%%%%%%%%%%%%%     Figure T2_Exp+Gau    %%%%%%%%%%%%%%%%%%%%%%%%%
	
	\begin{figure}[htbp]
		\includegraphics[width=\linewidth]{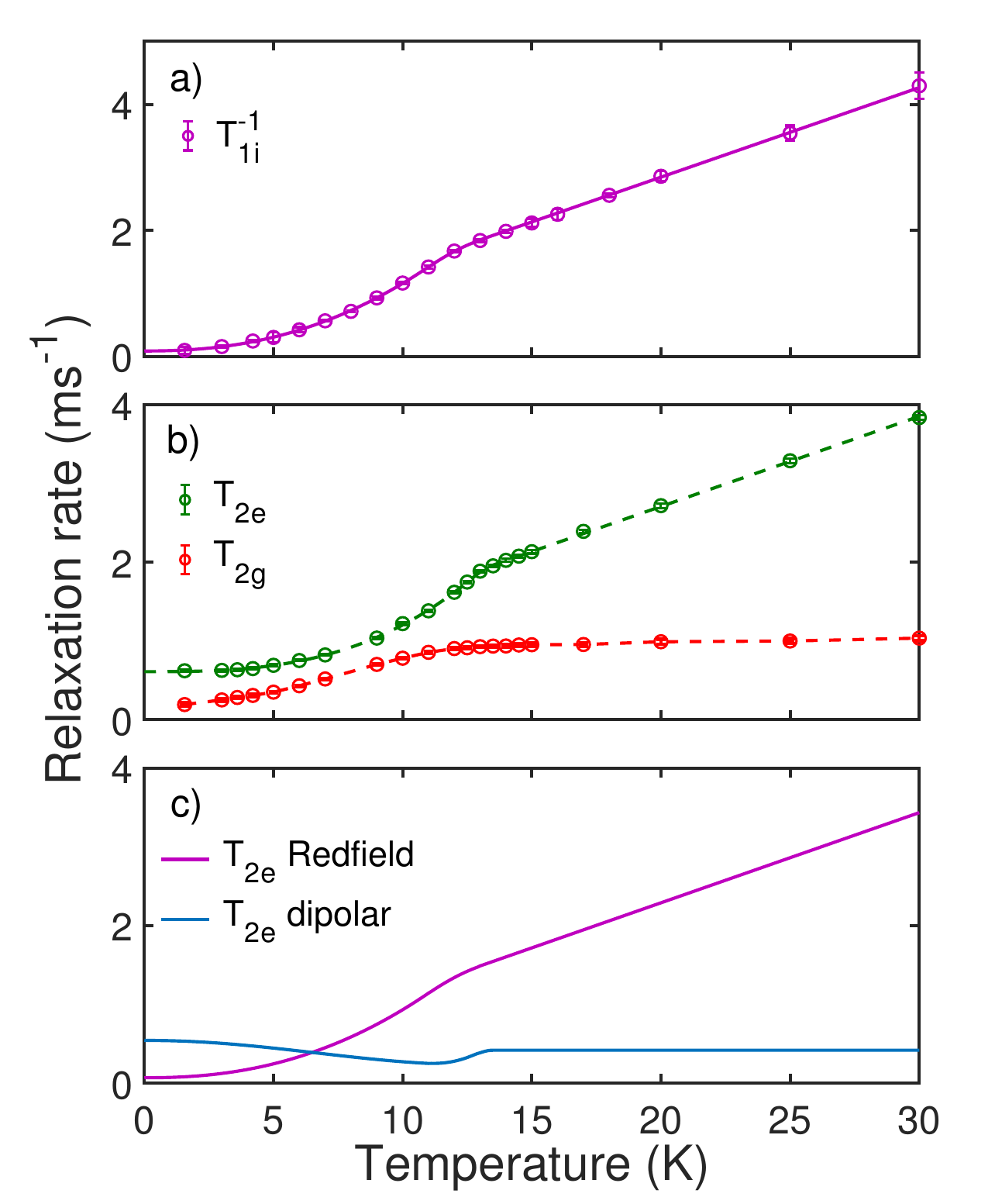}
		\caption{\label{T2_Exp+Gau} Temperature dependence of $^{93}$Nb transverse relaxation rate of the Nb$_3$Sn powder at 9.43\,T from Sample 2. Data are shown as dots with error bars less than dot size. The green and red data are the Lorentzian and Gaussian components of $T_2^{-1}$ respectively, achieved from fitting the recovery curve with Eq.~\ref{Exp+Gau}. The Lorentzian component is further separated into a Redfield component (purple curve) and a dipolar component (blue  curve). The Redfield component was calculated from the measurement of $T_{1i}$.}
	\end{figure}
	
	%\begin{figure}[htbp]
%		\includegraphics[width=\linewidth]{T2_E+G.pdf}
%		\caption{\label{T2_backup} This is for back up. Please choose the figure you prefer and delete %the rest one.}
%	\end{figure}
	
	 With our larger Sample 2 we were able to investigate the  behavior of $T_2^{-1}$ in considerable detail.  First and foremost,  there is no evidence of vortex dynamics in $T_{2meas}^{-1}$ such as was the case for Sample 1, Fig.~\ref{T2}.  This indicates that vortices are strongly pinned in the stoichiometric compound, at least on the kHz time scale consistent with its higher critical temperature, a behavior  which we will investigate  in future work. \red{The Redfield correction to $T_{2meas}^{-1}$ entails only $T_{1i}^{-1}$, the early time behavior of Eq.\ref{Master}, Fig.~\ref{T2_Exp+Gau}(a). The significant enhancement in the signal-to-noise ratio for our larger sample,  allows us to resolve Lorentzian and Gaussian  components of   $T_{2}^{-1}$ in the relaxation M(t), Eq.\ref{Exp+Gau},~\cite{Bac.98}	}
	
	\begin{equation}
		M(t) = M_0 e^{-\frac{t}{T_{2e}}} e^{-(\frac{t}{T_{2g}})^2},
		\label{Exp+Gau}
	\end{equation}
	
     	\noindent where $T_{2e}$ (green data) and  $T_{2g}$ (red data) are the Lorentzian and Gaussian components of $T_{2meas}^{-1}$ respectively, Fig.~\ref{T2_Exp+Gau}(b). We emphasize that for these data, the contribution to \blue{$T_{2e}$ from the Redfield  process was  independently determined from $T_{1i}^{-1}$, Eq.~\ref{Redfield2}, as $T_{2e \cdot Redfield}$ shown by the purple curve in Fig.~\ref{T2_Exp+Gau}(c)) with $\kappa=0.8$. The latter comes from the requirement that the Redfield contribution go to zero at zero temperature. $T_{2e}$ is then separated into $T_{2e \cdot Redfield}$  and $T_{2e \cdot dipolar}$, Fig.~\ref{T2_Exp+Gau}(c).} 
	   	
    	\blue{
    	\begin{equation}
    		T_{2e}^{-1} = T_{2e \cdot dipolar}^{-1} + T_{2e \cdot Redfield}^{-1} = T_{2e \cdot dipolar}^{-1} + \kappa \cdot T_{1i}^{-1}
    		\label{Redfield2}
    	\end{equation}	
    	}

    	In the normal state \blue{$T_{2g}$ (red curve in Fig.~\ref{T2_Exp+Gau}(b)}, was found to be  temperature independent as expected for the dipole-dipole interaction. However, what is unusual is our discovery of a temperature independent Lorentzian contribution to the dipolar $T_2^{-1}$ in the normal state which must be attributed to the indirect interaction. This was already evident in Fig.~\ref{T2} from the normal state extrapolations  to zero temperature where the lifetime of the nuclear spin state is  irrelevant.   We discuss the very interesting behavior in the superconducting state \red{later}. \blue{$T_{2g}$} is  temperature independent  in the normal state $\sim 1$\,ms$^{-1}$, and a factor of 20 smaller than what we calculated based on the square root of the second moment of the \red{Nb nuclear spin} dipolar field in Nb$_3$Sn, with the following relation~\cite{Abr.61}:
	
	\begin{equation}
		\overline{\Delta \omega ^2} = M_2 = (\frac{\mu_0}{4\pi})^2 \frac{3}{4} \gamma ^4 \hbar ^2 I (I+1) \sum_{k} \frac{(1-3\cos ^2 \theta)^2}{r_{jk}^6},
		\label{Dipole}
	\end{equation}
	
	\noindent where $\gamma$ is the gyromagnetic ratio, ${\bf r}_{jk}$ is the separation vector of two nuclei, $j$ and $k$, at an angle $\theta$ to the magnetic field, summed over all nuclei $k$ in the Nb$_3$Sn structure. As was mentioned in the introduction, this is evidence of a very strong, indirect exchange interaction between Nb nuclei~\cite{Sli.13} \red{that} reduces the dipolar $T_2^{-1}$, \red{substantially} decreasing the homogeneous contribution to the spectral NMR lineshape.
	
	In the superconducting state, the Redfield component is gapped out in the low temperature limit following the behavior of $T_{1i}$,  Fig.~\ref{T2_Exp+Gau}(a).  The dipolar component remains non-zero in the low temperature limit, \red{ still an order of magnitude less than the direct dipole value, consistent with the results of} Sample 1; \red{however, predominantly of Lorentzian character.} The Gaussian component decreases  with temperature toward zero,  apparently reflecting \red{the} decrease in quasiparticle density. We also attribute this to a modification of the indirect exchange interaction in the superconducting state.  This new result \red{is} important for understanding the interplay between superconductivity and magnetic RKKY interactions in metallic systems.

	\section{Conclusion}
	In summary, the superconducting order parameter amplitude is strongly suppressed by magnetic field in Nb$_3$Sn as determined from NMR longitudinal relaxation measurements.  From transverse relaxation we found an anomalously high nuclear spin coherence, which we attribute to a strong indirect RKKY exchange interaction between Nb nuclei. We discovered that this interaction has both Gaussian and Lorentzian character \red{that} is modified in the superconducting state to \red{become} mainly Lorentzian. We also identified vortex dynamics and vortex pinning that is sample dependent.
	
	\section{Acknowledgement}
	We thank Jim Sauls and Andy Mounce for  helpful discussions and Michael Brown for providing Sample 1.  We also acknowledge assistance from John Scott and Daehan Park in the ultra-low temperature laboratory. This work was supported by the U.S. Department of Energy, Office of Science, National Quantum Information  Science Research Centers, Superconducting Quantum Materials and Systems Center (SQMS) under contract No. DE-AC02-07CH11359. A portion of this work was performed at the National High Magnetic Field Laboratory, which is supported by National Science Foundation Cooperative Agreement No. DMR-2128556 and the State of Florida. The research work at Florida State University was primarily funded by the US. Department of Energy, Office of High Energy Physics, DE-SC0012083.

\newpage
\section{SUPPLEMENTARY MATERIALS}

\section{NMR Spectrum}
\begin{figure}[ht]
	\includegraphics[width=\linewidth]{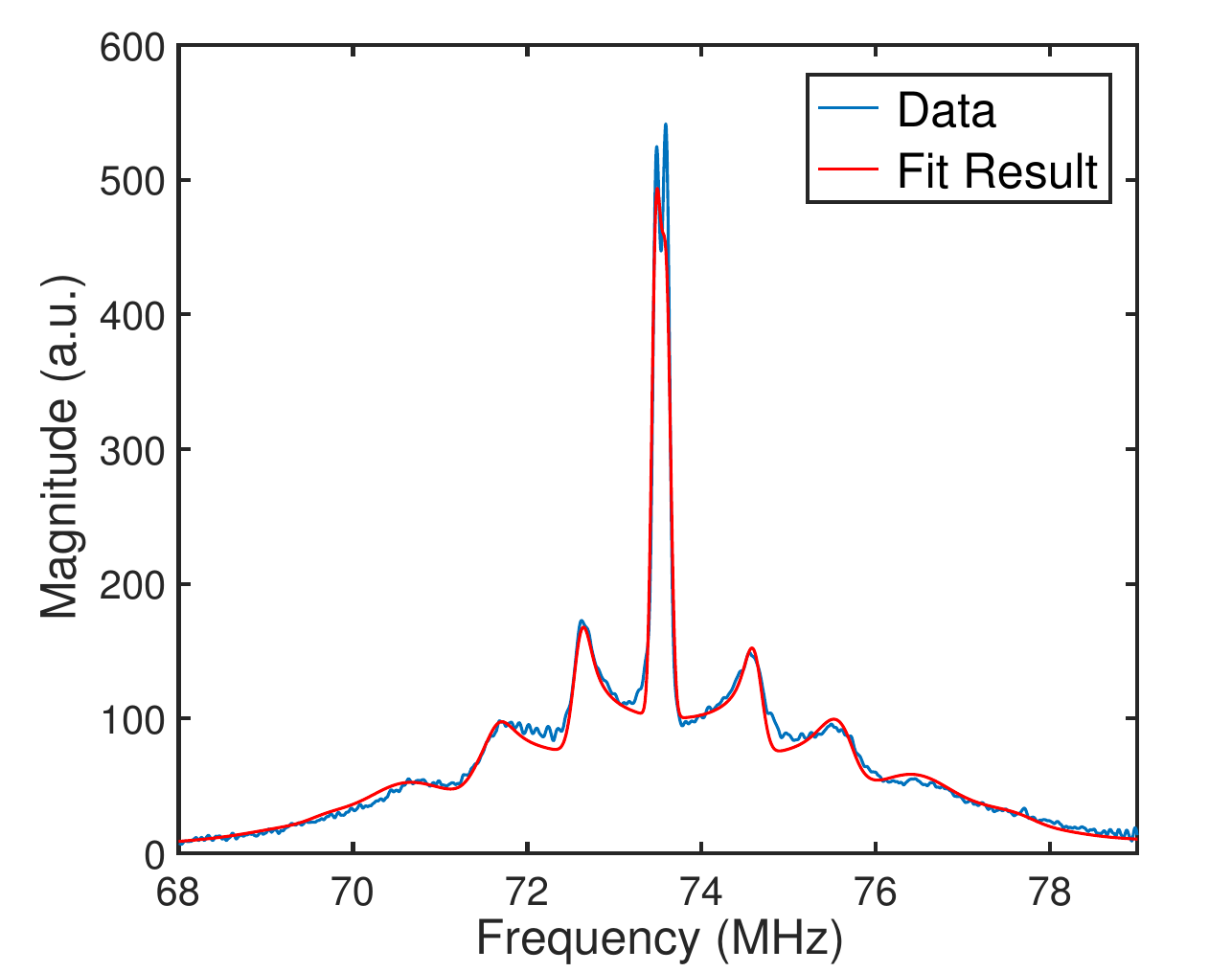}
	\label{Spectrum}
	\caption{NMR $^{93}$Nb spectrum of Sample 1.  The field sweep spectrum showing raw data in blue and fitted to the theory for random oriented powder in red; seven components of the nine are evident where two of the satellites at highest and lowest frequencies are unresolved. The central component is evident with a resolved splitting as expected in second order perturbation theory in the electric field gradient. Similar spectra were measured for Sample 2. }
\end{figure}

\newpage
\section{Critical Field}

\begin{figure}[ht]
	\includegraphics[width=\linewidth]{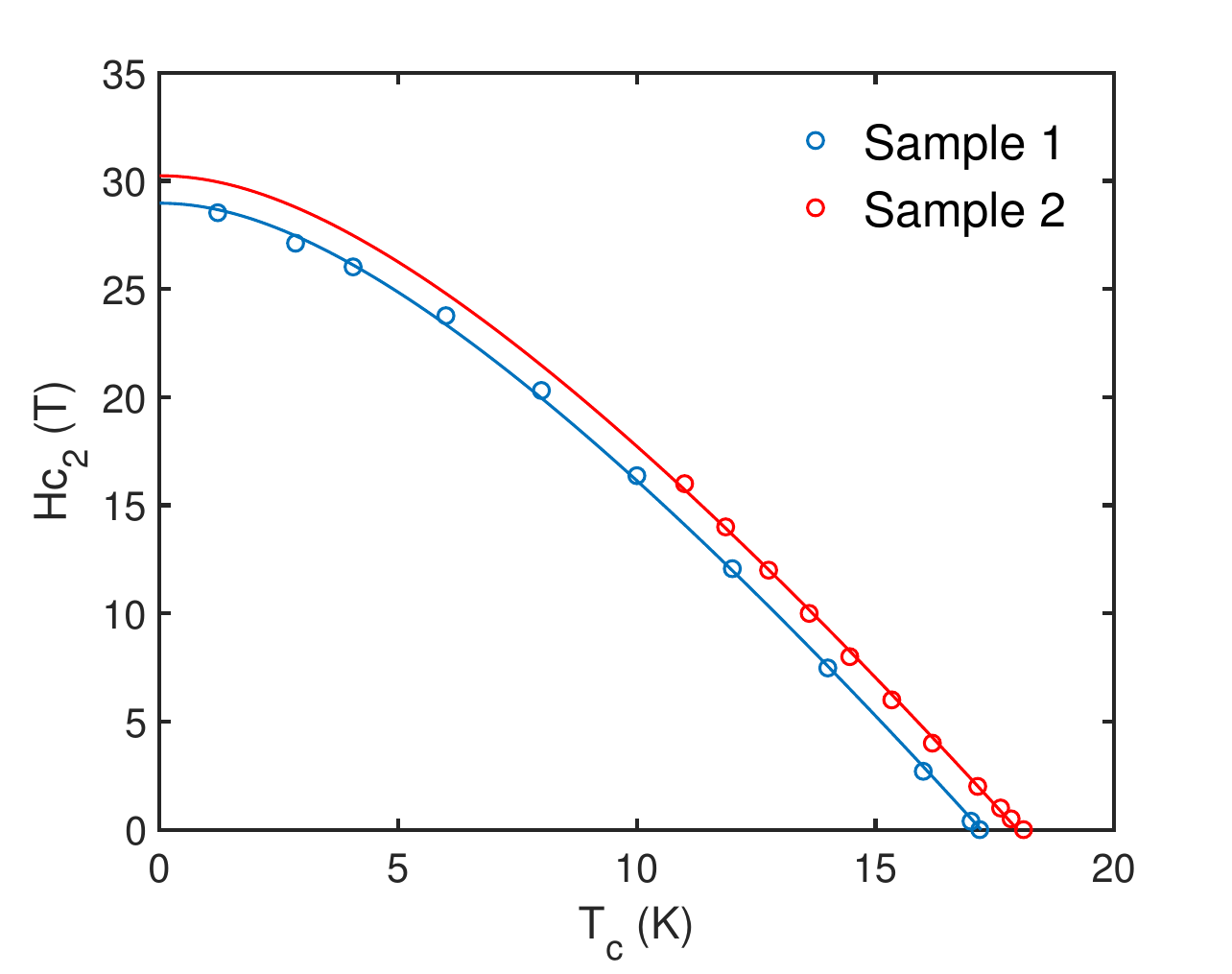}
	\label{Hc2}
	\caption{Temperature dependence of the upper critical field of Nb$_3$Sn. The upper critical field are evaluated with a 90\%Rn criterion from transport measurements. The blue curve (Sample 1) and red curve (Sample 2) are calculated using Werthamer, Helfand and Hohenberg (WHH) fitting.}
\end{figure}

\newpage
\section{SEM Image}
\begin{figure}[ht]
	\includegraphics[width=\linewidth]{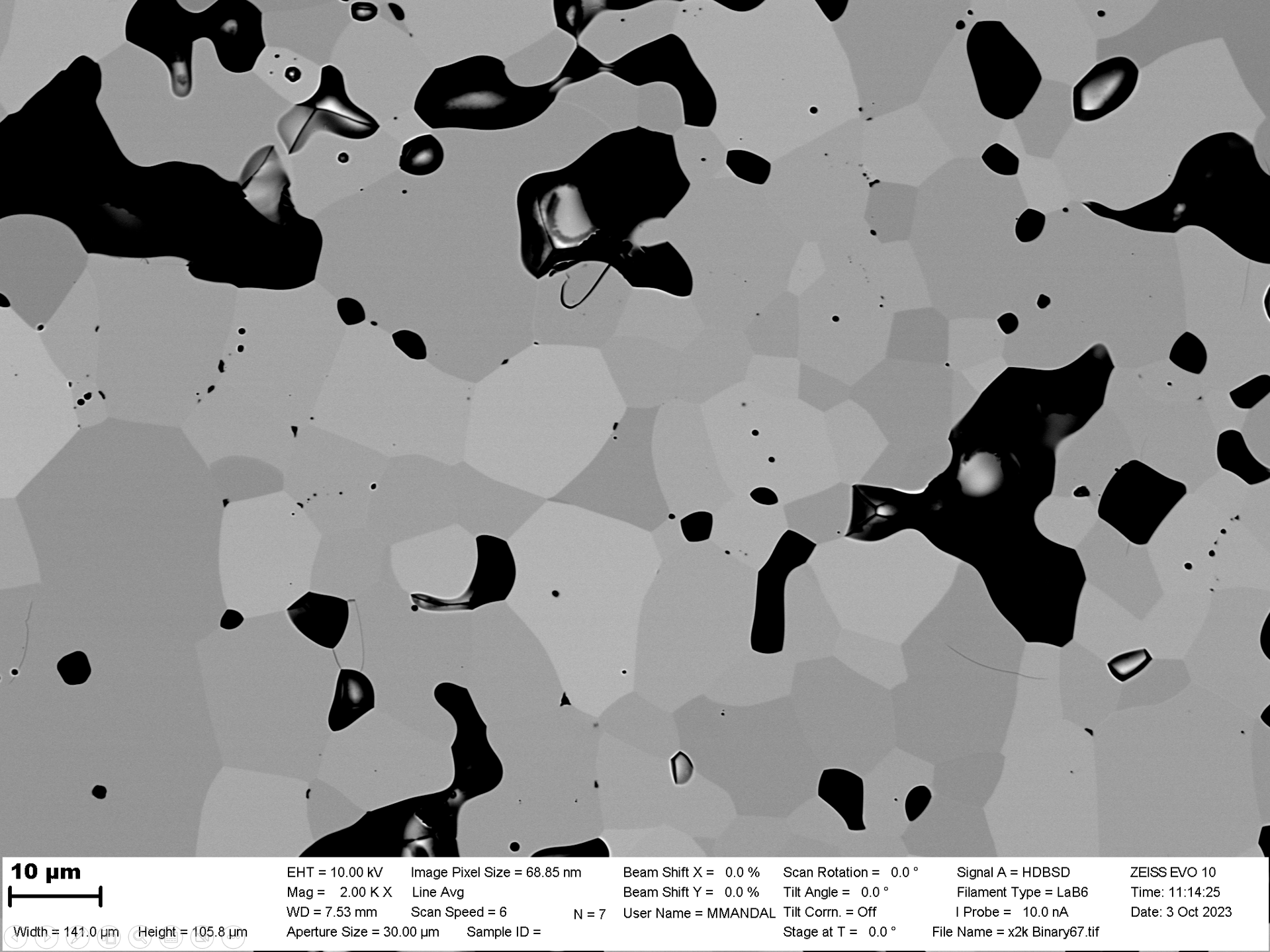}
	\label{SEM}
	\caption{SEM image of Sample 2, prior to ball milling. Note the uniform shade of what appear to be single crystal grains.}
\end{figure}

\newpage
\section{Resistance Measurement}
\begin{figure}[ht]
	\includegraphics[width=\linewidth]{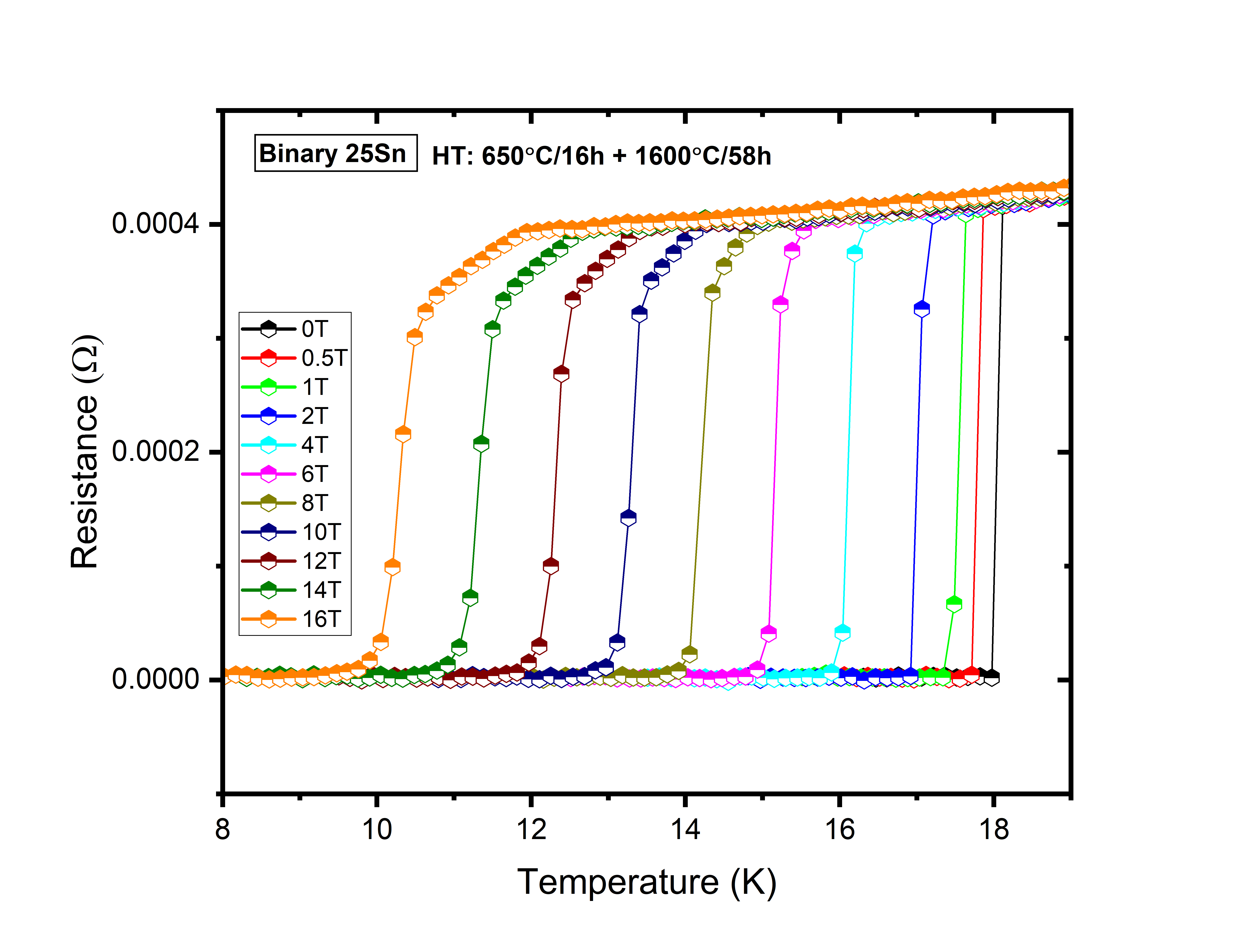}
	\label{Resistance}
	\caption{Temperature dependence of the resistance at different magnetic fields for Sample 2.}
\end{figure}

\newpage	
	\bibliography{Nb3Sn_Paper2_Full}

\begin{thebibliography}{17}%
\makeatletter
\providecommand \@ifxundefined [1]{%
 \@ifx{#1\undefined}
}%
\providecommand \@ifnum [1]{%
 \ifnum #1\expandafter \@firstoftwo
 \else \expandafter \@secondoftwo
 \fi
}%
\providecommand \@ifx [1]{%
 \ifx #1\expandafter \@firstoftwo
 \else \expandafter \@secondoftwo
 \fi
}%
\providecommand \natexlab [1]{#1}%
\providecommand \enquote  [1]{``#1''}%
\providecommand \bibnamefont  [1]{#1}%
\providecommand \bibfnamefont [1]{#1}%
\providecommand \citenamefont [1]{#1}%
\providecommand \href@noop [0]{\@secondoftwo}%
\providecommand \href [0]{\begingroup \@sanitize@url \@href}%
\providecommand \@href[1]{\@@startlink{#1}\@@href}%
\providecommand \@@href[1]{\endgroup#1\@@endlink}%
\providecommand \@sanitize@url [0]{\catcode `\\12\catcode `\$12\catcode
  `\&12\catcode `\#12\catcode `\^12\catcode `\_12\catcode `\%12\relax}%
\providecommand \@@startlink[1]{}%
\providecommand \@@endlink[0]{}%
\providecommand \url  [0]{\begingroup\@sanitize@url \@url }%
\providecommand \@url [1]{\endgroup\@href {#1}{\urlprefix }}%
\providecommand \urlprefix  [0]{URL }%
\providecommand \Eprint [0]{\href }%
\providecommand \doibase [0]{https://doi.org/}%
\providecommand \selectlanguage [0]{\@gobble}%
\providecommand \bibinfo  [0]{\@secondoftwo}%
\providecommand \bibfield  [0]{\@secondoftwo}%
\providecommand \translation [1]{[#1]}%
\providecommand \BibitemOpen [0]{}%
\providecommand \bibitemStop [0]{}%
\providecommand \bibitemNoStop [0]{.\EOS\space}%
\providecommand \EOS [0]{\spacefactor3000\relax}%
\providecommand \BibitemShut  [1]{\csname bibitem#1\endcsname}%
\let\auto@bib@innerbib\@empty
%</preamble>
\bibitem [{\citenamefont {Zhou}\ \emph {et~al.}(2011)\citenamefont {Zhou},
  \citenamefont {Jo}, \citenamefont {Hawn~Sung}, \citenamefont {Zhou},
  \citenamefont {Lee},\ and\ \citenamefont {Larbalestier}}]{Zho.11b}%
  \BibitemOpen
  \bibfield  {author} {\bibinfo {author} {\bibfnamefont {J.}~\bibnamefont
  {Zhou}}, \bibinfo {author} {\bibfnamefont {Y.}~\bibnamefont {Jo}}, \bibinfo
  {author} {\bibfnamefont {Z.}~\bibnamefont {Hawn~Sung}}, \bibinfo {author}
  {\bibfnamefont {H.}~\bibnamefont {Zhou}}, \bibinfo {author} {\bibfnamefont
  {P.~J.}\ \bibnamefont {Lee}},\ and\ \bibinfo {author} {\bibfnamefont {D.~C.}\
  \bibnamefont {Larbalestier}},\ }\href {https://doi.org/10.1063/1.3643055}
  {\bibfield  {journal} {\bibinfo  {journal} {Applied Physics Letters}\
  }\textbf {\bibinfo {volume} {99}},\ \bibinfo {pages} {122507} (\bibinfo
  {year} {2011})}\BibitemShut {NoStop}%
\bibitem [{\citenamefont {Zhou}(2011)}]{Zho.11}%
  \BibitemOpen
  \bibfield  {author} {\bibinfo {author} {\bibfnamefont {J.}~\bibnamefont
  {Zhou}},\ }\href@noop {} {\emph {\bibinfo {title} {The effects of variable
  tin content on the properties of A15 superconducting niobium-3-tin}}}\
  (\bibinfo  {publisher} {The Florida State University},\ \bibinfo {year}
  {2011})\BibitemShut {NoStop}%
\bibitem [{\citenamefont {Mandal}\ \emph {et~al.}(2025)\citenamefont {Mandal},
  \citenamefont {Tarantini}, \citenamefont {Starch}, \citenamefont {Lee},\ and\
  \citenamefont {Larbalestier}}]{Man.25}%
  \BibitemOpen
  \bibfield  {author} {\bibinfo {author} {\bibfnamefont {M.}~\bibnamefont
  {Mandal}}, \bibinfo {author} {\bibfnamefont {C.}~\bibnamefont {Tarantini}},
  \bibinfo {author} {\bibfnamefont {W.~L.}\ \bibnamefont {Starch}}, \bibinfo
  {author} {\bibfnamefont {P.~J.}\ \bibnamefont {Lee}},\ and\ \bibinfo {author}
  {\bibfnamefont {D.~C.}\ \bibnamefont {Larbalestier}},\ }\href
  {https://doi.org/10.1109/TASC.2024.3509388} {\bibfield  {journal} {\bibinfo
  {journal} {IEEE Transactions on Applied Superconductivity}\ }\textbf
  {\bibinfo {volume} {35}},\ \bibinfo {pages} {1} (\bibinfo {year}
  {2025})}\BibitemShut {NoStop}%
\bibitem [{\citenamefont {Zhai}\ \emph {et~al.}(2024)\citenamefont {Zhai},
  \citenamefont {Halperin}, \citenamefont {Reyes}, \citenamefont {Posen},
  \citenamefont {Sung}, \citenamefont {Tarantini}, \citenamefont {Brown},\ and\
  \citenamefont {Larbalestier}}]{Gan.24}%
  \BibitemOpen
  \bibfield  {author} {\bibinfo {author} {\bibfnamefont {G.}~\bibnamefont
  {Zhai}}, \bibinfo {author} {\bibfnamefont {W.}~\bibnamefont {Halperin}},
  \bibinfo {author} {\bibfnamefont {A.}~\bibnamefont {Reyes}}, \bibinfo
  {author} {\bibfnamefont {S.}~\bibnamefont {Posen}}, \bibinfo {author}
  {\bibfnamefont {Z.}~\bibnamefont {Sung}}, \bibinfo {author} {\bibfnamefont
  {C.}~\bibnamefont {Tarantini}}, \bibinfo {author} {\bibfnamefont
  {M.}~\bibnamefont {Brown}},\ and\ \bibinfo {author} {\bibfnamefont
  {D.}~\bibnamefont {Larbalestier}},\ }\href
  {https://doi.org/10.1088/1361-6668/ad5fbf} {\bibfield  {journal} {\bibinfo
  {journal} {Superconductor Science and Technology}\ }\textbf {\bibinfo
  {volume} {37}},\ \bibinfo {pages} {085020} (\bibinfo {year}
  {2024})}\BibitemShut {NoStop}%
\bibitem [{\citenamefont {Recchia}\ \emph {et~al.}(1997)\citenamefont
  {Recchia}, \citenamefont {Martindale}, \citenamefont {Pennington},
  \citenamefont {Hults},\ and\ \citenamefont {Smith}}]{Rec.97}%
  \BibitemOpen
  \bibfield  {author} {\bibinfo {author} {\bibfnamefont {C.~H.}\ \bibnamefont
  {Recchia}}, \bibinfo {author} {\bibfnamefont {J.~A.}\ \bibnamefont
  {Martindale}}, \bibinfo {author} {\bibfnamefont {C.~H.}\ \bibnamefont
  {Pennington}}, \bibinfo {author} {\bibfnamefont {W.~L.}\ \bibnamefont
  {Hults}},\ and\ \bibinfo {author} {\bibfnamefont {J.~L.}\ \bibnamefont
  {Smith}},\ }\href {https://doi.org/10.1103/PhysRevLett.78.3543} {\bibfield
  {journal} {\bibinfo  {journal} {Phys. Rev. Lett.}\ }\textbf {\bibinfo
  {volume} {78}},\ \bibinfo {pages} {3543} (\bibinfo {year}
  {1997})}\BibitemShut {NoStop}%
\bibitem [{\citenamefont {Bachman}\ \emph {et~al.}(1998)\citenamefont
  {Bachman}, \citenamefont {Reyes}, \citenamefont {Mitrovic}, \citenamefont
  {Halperin}, \citenamefont {Kleinhammes}, \citenamefont {Kuhns},\ and\
  \citenamefont {Moulton}}]{Bac.98}%
  \BibitemOpen
  \bibfield  {author} {\bibinfo {author} {\bibfnamefont {H.~N.}\ \bibnamefont
  {Bachman}}, \bibinfo {author} {\bibfnamefont {A.~P.}\ \bibnamefont {Reyes}},
  \bibinfo {author} {\bibfnamefont {V.~F.}\ \bibnamefont {Mitrovic}}, \bibinfo
  {author} {\bibfnamefont {W.~P.}\ \bibnamefont {Halperin}}, \bibinfo {author}
  {\bibfnamefont {A.}~\bibnamefont {Kleinhammes}}, \bibinfo {author}
  {\bibfnamefont {P.}~\bibnamefont {Kuhns}},\ and\ \bibinfo {author}
  {\bibfnamefont {W.~G.}\ \bibnamefont {Moulton}},\ }\href
  {https://doi.org/10.1103/PhysRevLett.80.1726} {\bibfield  {journal} {\bibinfo
   {journal} {Phys. Rev. Lett.}\ }\textbf {\bibinfo {volume} {80}},\ \bibinfo
  {pages} {1726} (\bibinfo {year} {1998})}\BibitemShut {NoStop}%
\bibitem [{\citenamefont {Ruderman}\ and\ \citenamefont
  {Kittel}(1954)}]{Rud.54}%
  \BibitemOpen
  \bibfield  {author} {\bibinfo {author} {\bibfnamefont {M.~A.}\ \bibnamefont
  {Ruderman}}\ and\ \bibinfo {author} {\bibfnamefont {C.}~\bibnamefont
  {Kittel}},\ }\href {https://journals.aps.org/pr/pdf/10.1103/PhysRev.96.99}
  {\bibfield  {journal} {\bibinfo  {journal} {Rev. Mod. Phys.}\ }\textbf
  {\bibinfo {volume} {96}},\ \bibinfo {pages} {99} (\bibinfo {year}
  {1954})}\BibitemShut {NoStop}%
\bibitem [{\citenamefont {Anderson}\ and\ \citenamefont
  {Weiss}(1953)}]{And.53}%
  \BibitemOpen
  \bibfield  {author} {\bibinfo {author} {\bibfnamefont {P.~W.}\ \bibnamefont
  {Anderson}}\ and\ \bibinfo {author} {\bibfnamefont {P.~R.}\ \bibnamefont
  {Weiss}},\ }\href {https://doi.org/10.1103/RevModPhys.25.269} {\bibfield
  {journal} {\bibinfo  {journal} {Rev. Mod. Phys.}\ }\textbf {\bibinfo {volume}
  {25}},\ \bibinfo {pages} {269} (\bibinfo {year} {1953})}\BibitemShut
  {NoStop}%
\bibitem [{\citenamefont {Slichter}(2013)}]{Sli.13}%
  \BibitemOpen
  \bibfield  {author} {\bibinfo {author} {\bibfnamefont {C.~P.}\ \bibnamefont
  {Slichter}},\ }\href@noop {} {\emph {\bibinfo {title} {Principles of magnetic
  resonance}}},\ Vol.~\bibinfo {volume} {1}\ (\bibinfo  {publisher} {Springer
  Science \& Business Media},\ \bibinfo {year} {2013})\BibitemShut {NoStop}%
\bibitem [{\citenamefont {Walstedt}\ \emph {et~al.}(1962)\citenamefont
  {Walstedt}, \citenamefont {Dowley},\ and\ \citenamefont
  {Froideveaux}}]{Wal.62}%
  \BibitemOpen
  \bibfield  {author} {\bibinfo {author} {\bibfnamefont {R.}~\bibnamefont
  {Walstedt}}, \bibinfo {author} {\bibfnamefont {E.~L.}\ \bibnamefont {Dowley},
  \bibfnamefont {M.~E.and~Hahn}},\ and\ \bibinfo {author} {\bibfnamefont
  {C.}~\bibnamefont {Froideveaux}},\ }\href
  {https://journals.aps.org/prl/pdf/10.1103/PhysRevLett.8.406} {\bibfield
  {journal} {\bibinfo  {journal} {Phys. Rev. Lett.}\ }\textbf {\bibinfo
  {volume} {8}},\ \bibinfo {pages} {406} (\bibinfo {year} {1962})}\BibitemShut
  {NoStop}%
\bibitem [{\citenamefont {Hahn}(1950)}]{Hah.50}%
  \BibitemOpen
  \bibfield  {author} {\bibinfo {author} {\bibfnamefont {E.~L.}\ \bibnamefont
  {Hahn}},\ }\href@noop {} {\bibfield  {journal} {\bibinfo  {journal} {Physical
  Review}\ }\textbf {\bibinfo {volume} {80}},\ \bibinfo {pages} {580} (\bibinfo
  {year} {1950})}\BibitemShut {NoStop}%
\bibitem [{\citenamefont {Suter}\ \emph {et~al.}(1998)\citenamefont {Suter},
  \citenamefont {Mali}, \citenamefont {Roos},\ and\ \citenamefont
  {Brinkmann}}]{Sut.98}%
  \BibitemOpen
  \bibfield  {author} {\bibinfo {author} {\bibfnamefont {A.}~\bibnamefont
  {Suter}}, \bibinfo {author} {\bibfnamefont {M.}~\bibnamefont {Mali}},
  \bibinfo {author} {\bibfnamefont {J.}~\bibnamefont {Roos}},\ and\ \bibinfo
  {author} {\bibfnamefont {D.}~\bibnamefont {Brinkmann}},\ }\href
  {https://doi.org/10.1088/0953-8984/10/26/022} {\bibfield  {journal} {\bibinfo
   {journal} {Journal of Physics: Condensed Matter}\ }\textbf {\bibinfo
  {volume} {10}},\ \bibinfo {pages} {5977} (\bibinfo {year}
  {1998})}\BibitemShut {NoStop}%
\bibitem [{\citenamefont {Halperin}\ and\ \citenamefont
  {Varoquaux}(1990)}]{Hal.90}%
  \BibitemOpen
  \bibfield  {author} {\bibinfo {author} {\bibfnamefont {W.~P.}\ \bibnamefont
  {Halperin}}\ and\ \bibinfo {author} {\bibfnamefont {E.}~\bibnamefont
  {Varoquaux}},\ }in\ \href
  {https://doi.org/https://doi.org/10.1016/B978-0-444-87476-4.50013-3} {\emph
  {\bibinfo {booktitle} {Helium Three}}},\ \bibinfo {series} {Modern Problems
  in Condensed Matter Sciences}, Vol.~\bibinfo {volume} {26},\ \bibinfo
  {editor} {edited by\ \bibinfo {editor} {\bibfnamefont {W.~P.}\ \bibnamefont
  {Halperin}}\ and\ \bibinfo {editor} {\bibfnamefont {L.~P.}\ \bibnamefont
  {Pitaevskii}}}\ (\bibinfo  {publisher} {Elsevier},\ \bibinfo {year} {1990})\
  pp.\ \bibinfo {pages} {353--522}\BibitemShut {NoStop}%
\bibitem [{\citenamefont {Ding}\ \emph {et~al.}(2023)\citenamefont {Ding},
  \citenamefont {Zhao}, \citenamefont {Jiang}, \citenamefont {Wang},
  \citenamefont {Feng},\ and\ \citenamefont {Zhang}}]{Din.23}%
  \BibitemOpen
  \bibfield  {author} {\bibinfo {author} {\bibfnamefont {S.}~\bibnamefont
  {Ding}}, \bibinfo {author} {\bibfnamefont {D.}~\bibnamefont {Zhao}}, \bibinfo
  {author} {\bibfnamefont {T.}~\bibnamefont {Jiang}}, \bibinfo {author}
  {\bibfnamefont {H.}~\bibnamefont {Wang}}, \bibinfo {author} {\bibfnamefont
  {D.}~\bibnamefont {Feng}},\ and\ \bibinfo {author} {\bibfnamefont
  {T.}~\bibnamefont {Zhang}},\ }\href@noop {} {\bibfield  {journal} {\bibinfo
  {journal} {Quantum Frontiers}\ }\textbf {\bibinfo {volume} {2}},\ \bibinfo
  {pages} {3} (\bibinfo {year} {2023})}\BibitemShut {NoStop}%
\bibitem [{\citenamefont {Fetter}\ and\ \citenamefont
  {Hohenberg}(1968)}]{Fet.68}%
  \BibitemOpen
  \bibfield  {author} {\bibinfo {author} {\bibfnamefont {A.~L.}\ \bibnamefont
  {Fetter}}\ and\ \bibinfo {author} {\bibfnamefont {P.~C.}\ \bibnamefont
  {Hohenberg}},\ }in\ \href@noop {} {\emph {\bibinfo {booktitle}
  {Superconductivity}}},\ \bibinfo {editor} {edited by\ \bibinfo {editor}
  {\bibfnamefont {R.~D.}\ \bibnamefont {Parks}}}\ (\bibinfo  {publisher}
  {Dekker},\ \bibinfo {year} {1968})\ pp.\ \bibinfo {pages}
  {817--923}\BibitemShut {NoStop}%
\bibitem [{\citenamefont {Redfield}(1957)}]{Red.57}%
  \BibitemOpen
  \bibfield  {author} {\bibinfo {author} {\bibfnamefont {A.~G.}\ \bibnamefont
  {Redfield}},\ }\href {https://doi.org/10.1147/rd.11.0019} {\bibfield
  {journal} {\bibinfo  {journal} {IBM Journal of Research and Development}\
  }\textbf {\bibinfo {volume} {1}},\ \bibinfo {pages} {19} (\bibinfo {year}
  {1957})}\BibitemShut {NoStop}%
\bibitem [{\citenamefont {Abragam}(1961)}]{Abr.61}%
  \BibitemOpen
  \bibfield  {author} {\bibinfo {author} {\bibfnamefont {A.}~\bibnamefont
  {Abragam}},\ }\href {https://books.google.com/books?id=9M8U_JK7K54C} {\emph
  {\bibinfo {title} {The Principles of Nuclear Magnetism}}},\ International
  series of monographs on physics\ (\bibinfo  {publisher} {Clarendon Press},\
  \bibinfo {year} {1961})\BibitemShut {NoStop}%
\end{thebibliography}%
\end{document}